\documentclass[aps,prb,showpacs,twocolumn,superscriptaddress]{revtex4-1}
\usepackage{bm,color,amsmath,amssymb,mathrsfs,latexsym,graphicx,psfrag,mathtools}
\usepackage{color}
\usepackage[dvipsnames]{xcolor}
\usepackage[hidelinks,colorlinks,linkcolor=blue,
citecolor=blue,urlcolor=blue]{hyperref}
\usepackage{dsfont}

\usepackage{empheq}

\let\oldAA\AA
\renewcommand{\AA}{\text{\normalfont\oldAA}}

\usepackage{comment}

\begin{document}
\title{Comment on {\it Nontrivial Quantum Geometry 
        and the Strength of Electron-Phonon Coupling,\\
arXiv:2305.02340}, J. Yu, C. J. Ciccarino, R. Bianco, I. Errea, 
P. Narang, B. A. Bernevig} 

\author{Warren E. Pickett}
\affiliation{Department of Physics and Astronomy, University of California Davis,
   Davis CA 95616 USA}
\begin{abstract}
The titular manuscript invites a description of the background of 
research on MgB$_2$,
whose unique electron-phonon coupling strength and record superconducting
T$_c$ for such a material at ambient pressure was demonstrated by the
combined efforts of several groups two decades ago in terms of conventional
but highly developed Migdal-Eliashberg theory. This Comment provides some of
the basic features of the theoretical understanding of MgB$_2$ and draws
contrasts with the model of Yu {\it et al.}
\end{abstract}
\maketitle

\section{Background}
The theory of electron-phonon coupled (EPC) superconductivity has
progressed from its proposal in 1957 to mature {\it ab initio}
superconducting density functional theory (SCDFT)\cite{SCDFT} 
and a precise specification
of EPC\cite{Giustino} in the early
21st century, for materials without dominant electron-electron
interactions. This development, requiring perfection of electronic
band theory, then phonon spectrum calculations, and EPC matrix elements,
for input into the Eliashberg equations, has been chronicled in a
recent review.\cite{WEP2023} The discovery of superconductivity MgB$_2$ with
critical temperature T$_c$=40 K ($^{10}B$ isotope) led to a flurry of
both experimental and theoretical study of the mechanism and its 
properties.\cite{PhysicaC2007}
The result of many prompt theoretical studies
\cite{An2001,Kortus2001,Kong2001,Liu2001,Yildirim2001,Bohnen2001,
Choi2002,An2002,WEP2003,Choi2003,Mazin2003,Choi2006} 
within a year or two of the discovery was that the standard theory of
electron-phonon coupling (EPC) applied - conventional theory with
an exceptional result - and also provided deep insight into EPC and raised
possibilities of related materials,\cite{LiBC} quite possibly with higher critical
temperature T$_c$.

Yu and co-authors\cite{Yu2023} (Yu2023) have presented a model of electron-phonon
coupling (EPC) and superconducting T$_c$ in electron-phonon
coupled metals (specifically MgB$_2$), identifying
a contribution with topological character and providing expectation that
this formulation may provide guidance into discovery of higher T$_c$
superconductivity. An effective Hamiltonian  application to 
MgB$_2$ was provided. Given the considerable effort into design and
discovery of MgB$_2$-like materials -- strong covalent bonds doped 
to be metallic -- these claims merit a comparison with the established,
and highly successful, theory of the superconducting mechanism and
several properties    
of MgB$_2$ that was already in place in the 
early 2000s.\cite{An2001,Kortus2001,Kong2001,Liu2001,Yildirim2001,Bohnen2001,
Choi2002,An2002,WEP2003,Choi2003,Mazin2003,Choi2006} 

A basic way to approach this Comment is: does
this new picture contain the essential physics of MgB$_2$ 
superconductivity and T$_c$, which is the intention that is used 
to support other claims? This Comment begins with a synopsis of
the very rapid development of the conceptual understanding and broad
application to many properties of MgB$_2$ using conventional
theoretical knowledge and codes. The following section points out
several aspects of the topic of this Comment. The Summary lists
take-home items to consider. 

\section{The conventional theory applied to MgB$_2$}
Superconductivity theory of EPC metals is based\cite{WEP2023} 
on (i) density functional theory of
electronic structure and phonon spectra, (ii) electron-phonon coupling
based on self-consistent density response to provide the change in
electronic potential due to phonon (atomic) displacements,\cite{Giustino} 
and (iii) Migdal-Eliashberg
theory of the electron and phonon self-energies. Superconductivity is
emergent in this theory: T$_c$ is determined
as the temperature at which an anomalous electron 
self-energy (superconducting gap)
appears. The essential components for MgB$_2$ are as follows.

{\it The electronic spectrum} contains (a) two bands crossing the
Fermi level E$_F$ (also referred to as the chemical potential $\mu$)
near $\Gamma$, arising from strong in-plane B-B $\sigma$ bonds giving
rise to two (effectively identical) Fermi cylinders centered along 
$\Gamma$-$A$, containing within the Fermi wavevector $k_F$ about 3\% of
the Brillouin zone volume ($k_F$ is 1/6 of the average radius $k_{bz}$ of the
hexagonal BZ, thus containing 3\% of the volume), and 
(b) larger Fermi surfaces (FSs) arising from
the out-of-plane $p_z$ ($\pi$) orbitals whose shapes are of little
interest.\cite{An2001,Kortus2001,Kong2001,Liu2001,Yildirim2001,Bohnen2001,
Choi2002,An2002,WEP2003,Choi2003,Mazin2003,Choi2006} 
Mg has given up its valence electrons to the honeycomb structure
B layers, giving a graphene-like band structure except for a Madelung
potential shifts that create a natural hole-doping of the $\sigma$ bands
by 0.11 holes/f.u.\cite{Deepa} [Note: 2D language for $k$-space quantities
is used here and below.]

{\it The phonon spectrum} is unique and provides the key. 
Aside from some Mg mixing
at low frequencies, it
is roughly graphene-like except for an extremely sharp and strong Kohn
anomaly at $q$=$2k_F$, which is the upper limit at which electrons can be
scattered intra-FS from cylinder to cylinder. 
The in-plane bond-stretch $E_g$ pair of modes lie $\sim$30\% lower in energy than the
unrenormalized modes closer to the zone boundary; these modes are extremely
broad, discussed further below.
These modes are thereby extremely strongly coupled\cite{An2001,WEP2003} -- mode
$\lambda_{q,E_g}$$\sim$ 20-25 -- while all other phonons are weakly coupled
($\lambda_{q,\nu}$$\sim$ 0.3-0.4 or less; there are no $\sigma-\sigma$ 
scattering processes for $q>2k_F$).
The $E_g$ modes are exceptional because B-B bonding states are so
strongly coupled to the $E_g$ stretch and angle-bending displacements 
with their
very large density response, leading to an unexpectedly large 
kernel of the electron-phonon matrix element.

{\it Electron-phonon coupling} in MgB$_2$  is challenging 
to calculate accurately.
The neutron spectroscopy based expression of Allen,\cite{Allen1972,PBA-MLC_1972} 
giving the contribution of each phonon to EPC strength $\lambda$,
is evaluated as an average over all mode $\lambda_{q\nu}$:
\begin{eqnarray}
\lambda = \frac{1}{3N_{at}}\sum_{q\nu}\lambda_{q\nu};~~~
    \lambda_{q\nu}  = \frac{2}{\pi N(\mu)}
             \frac{\gamma_{q\nu}}{\omega_{q\nu}^2},
\label{eqn:lambda}
\end{eqnarray}
the expression for the linewidth/frequency ratio
 $\gamma_{q\nu}/\omega_{q\nu}$ is given below. The extra factor of
$\omega_{q\nu}$ is from weighting to determine $\lambda$.

The strong and sharp 2D Kohn anomaly at $q$$=$$2k_F$ divides vanishing phonon 
linewidths $\gamma_{q,E_g}$=0 for $q$$>$$2k_F$ (no intra-cylinder scattering
processes) to very large and strongly $q$-dependent values  inside $2k_F$.
The coupling strength $\lambda_{q,E_g}$ is further enhanced
by the strong renormalization of the frequency $\omega_{q,E_g}$
below $q=2k_F$. It is revealing that because theory provides the value of
$\omega^2_{q\nu}$, renormalization downward is to around 2/3 of the value
without EPC (viz. isostructural AlB$_2$ without $\sigma$ Fermi surfaces, 
or $q$$>$$2k_F$ for the $E_g$ modes of MgB$_2$). Thus$\omega_{E_g}^2$,
which is what is given by phonon theory,
is renormalized to less than half of the values for $q$$>$$2k_F$,
MgB$_2$ is not so far from the looming structural instability of $q$=0
bond-stretching distortions (see {\it e.g.} Refs.~[\onlinecite
{WEP-Brazil,WEP-Design,WEP-Break}]). The still rather high phonon energy
provides a promising energy/temperature scale for T$_c$, and leaves some 
room for materials engineering toward stronger coupling. 
 
This discontinuity in $\gamma_{q,E_g}$ and sharp change in $\omega_{q,E_g}$
across $q$=$2k_F$ is highly
challenging to accommodate numerically with discrete $q$-mesh plus Fourier
interpolation methods assuming smoothness, largely accounting for variation 
amongst reported calculations.
Given the small region around $\Gamma$ that is so highly renormalized by
EPC, treating the $E_g$ matrix elements as constant in this region is justified.
Then\cite{Allen1972,Giustino2}
\begin{eqnarray}
\gamma_{q,E_g} &=& \frac{\pi}{\hbar} \sum_{k,n,n'} |M_{k,n;k+Q,n'}|^2 \nonumber \\ 
 & & \times \delta(E_{k,n}-\mu)
            \delta(E_{k+q,n'}-E_{k,n}-\hbar\omega_{q\nu}) \nonumber \\
 &\rightarrow& 4\pi \omega_{q\nu}  {\bar{M}_{E_g}^2} 
      \sum_k \delta(E_k-\mu)\delta(E_{k+q}-\mu)
     \nonumber \\
  & =&\frac{4\pi \omega_{q\nu}}{N(\mu)} \bar{M}^2_{E_g} \xi(q),
\label{eqn:gamma}
\end{eqnarray} 
where is $\xi(q)$ is the nesting function and factor of four takes care of the
scattering between the effectively identical pair of cylindrical FSs.
 
The phase space $\xi(q)$ for scattering across the cylindrical 
Fermi surfaces in the
integral for the linewidths, reducing to circular FSs in 2D,
is readily evaluated analytically,\cite{WEP-Brazil,WEP-Design,WEP-Break} obtaining
\begin{eqnarray}
\gamma_{q,E_g} &=& 4\pi\omega_{q,E_g} \bar{M_{E_g}^2} 
       \frac{N(\mu)}{ \sqrt{\eta_q (1-\eta_q^{2})^{1/2}  } },~~~\eta_q<1,
\end{eqnarray}
and zero for larger $q$; here $\eta_q$$\equiv$$q/2k_F$.
This expression reveals a divergence in $\gamma_{q,E_g}$
for $q$$\rightarrow$0 and for $q$$\rightarrow$$2k_F$ ($\omega_{q,E_g}$ is 
constant for $q$$<$$2k_F$, losing its renormalization as $\eta_q$ increases 
beyond unity). Both divergences are integrable, but
this strong $q$ variation, discontinuity at $2k_F$, and small area 
of integration makes (as already mentioned) $q$-mesh numerical methods based on smooth
functions problematic, and  
convergence of numerical sampling methods likely has never been achieved
for MgB$_2$. The highest resolution of this Kohn anomaly was shown for the
isostructural, isoelectronic Li$_x$BC in Ref.~[\onlinecite{WEP-Break}].
The progression in Eq.~(\ref{eqn:gamma}) was applied\cite{An2001} in an
evaluation of the strength of $\sigma$-$\sigma$ processes, treating the radial
$q$-dependence. The analytic behaviors were presented in Ref.~(\onlinecite{WEP-Break}).
Fortunately, very precise numbers are not the point; 
recognizing fully the underlying physics should be the objective.

{\it A summary of the EPC and T$_c$} in MgB$_2$ is this. \\
(i) Similar to B-doped diamond but discovered earlier, 
quasi-2D MgB$_2$ superconducts at 40K (isotope $^{10}B$)
due to strong EPC of electronic states in the B-B bonds to 
primarily bond stretch
phonon modes. The McMillan-Eliashberg spectral function $\alpha^2F(\omega)$
is dominated by this contribution, as shown in several papers.\\
(ii) The important Fermi surfaces are small and simply shaped, a pair of them being
cylindrical around the $(0,0,q_z)$ $\Gamma$-$A$ line, with 0.11 hole doping per f.u. 
The FSs have been studied experimentally, viz. in Ref.~[\onlinecite{Fletcher2004}].\\
(iii) Less than 3\% [$(2k_F/k_{bz})^2$=12\% $\times$ 2/9 of the branches]
of the phonons of MgB$_2$ 
provide\cite{An2001} a large majority of the 
coupling strength and T$_c$. Calculations of the
contribution of $\pi$-$\pi$, $\sigma$-$\pi,$ and $\pi$-$\sigma$ of the remaining phonons
(97\% of them) nevertheless account for 10-15K
of the final value of T$_c$ -- if they are considered as coming on top of the 
crucial $\sigma$-$\sigma$ coupling.\cite{Liu2001,Choi2003,Mazin2003,Choi2006} 
By these scatterings alone, MgB$_2$ would be a weakly-coupled  
non-superconductor.\\
(iv) As a result of the very strong coupling, the width of the $E_g$ phonon 
is around 40\% of
its frequency: once excited, this phonon undergoes only $\sim$three vibrational periods
before decaying into electron-hole pairs or multiple phonons. 
This huge width has been verified by
inelastic x-ray scattering\cite{Shukla2003,Baron2004} and superconducting gap
was quantified by Raman 
spectroscopy.\cite{Quilty2002} It should be noted that the experimental
width will include anharmonic and non-linear coupling contributions, both
involving electron -- multi-phonon scattering contributions, as discovered by
a few groups.\cite{Kortus2001,Liu2001,Yildirim2001}\\ 
(v) MgB$_2$ is calculated to be a two-gap (``two band'') 
superconductor,\cite{Liu2001,Choi2003,Mazin2003,Choi2006} 
large on the $\sigma$ FSs (65 meV) and small on the (larger) $\pi$ FSs
(15 meV). The predicted two gaps have been verified by
heat capacity $C_p(T)$ measurements below T$_c$. Several other experimental 
verifications of calculated properties have been verified. 
The experimental value of T$_c$
is approached closely only when the two-gap nature is incorporated in the theory. \\
(vi) Despite two decades of designs of higher T$_c$ versions of the MgB$_2$ paradigm,
involving handfuls of suggestions, no better such superconductor has been confirmed
experimentally. After two decades, MgB$_2$ remains a class unto itself.

\section{The model of Yu {\it et al.}} 
This model applies a parametrized two-center tight-binding  
model of the electronic spectrum 
to provide a new viewpoint on EPC as applied 
MgB$_2$. The objective is to identify and isolate a topological 
aspect to the EPC, to try to relate it
quantitatively to T$_c$, and to suggest that this viewpoint may provide direction toward
higher T$_c$ materials. The main points follow.

{\it The tight-binding electronic structure.} Certain parts of the results appear
to involve a 17 parameter Hamiltonian including Mg (Appendix H). Most results are
described in terms of a simpler effective tight-binding (TB) model of two 
$\sigma$ bonding bands (initially indicated as only one crossing $E_F$,
but becoming degenerate in their first-order-in-$k_{\Vert}$ simplification)
and two $\pi$ ($p_z$) bands, which must be very similar to the TB
parametrization of Refs.~[\onlinecite{Liu2001,Mazin2003}] with $t_{\sigma}$=2.5 eV,
$t_{\pi}$=1.5 eV. 
The curvature of the $\sigma$ bands 
away from $\Gamma$ (downward), given by the hopping parameter,  
is adopted from DFT calculations, and the position of the 
Fermi level (denoted $\mu \equiv E_F$) is also taken from DFT results. 
These choices lead to cylindrical FSs
effectively equivalent to the DFT surfaces. 3D dispersion of the $\pi$ bands
is necessary when their FSs are to be sampled.

{\it The phonon formalism}, and distinction between the $E_g$ pair and other phonons,
is discussed with substantial detail and sublattice indices in Appendices. However, phonons
are not explicitly used to evaluate $\lambda$. The only frequency that appears
is the McMillan $<$$\omega^2$$>$ factor in the denominator of
their expression for $\lambda$ is adapted from DFT and experiment, 
and is discussed below. This omission misses the essence of MgB$_2$.
   
{\it Electron-lattice coupling} is treated in the analytically tractable manner 
of supposing that the hopping integrals can be approximated as analytic 
(Gaussian approximation [GA]) functions $t(R)$ of atomic separation $R$: 
\begin{eqnarray}
t(R)=t_o e^{-\gamma R^2/2}. 
\end{eqnarray}
The form proposed in Harrison's classic discussion 
of TB modeling is, formally, for $p$-$p$ ($\ell=\ell'$=1) overlaps,
\begin{eqnarray}
t(R)&=& <\phi_p(0)|[\frac{p^2}{2m}+V(r)]|\phi_p'(R)> \nonumber \\
    &=&\propto \frac{1}{R^{\ell+\ell' +1}} = \frac{1}{R^3}.
\end{eqnarray}
Other choices beyond Harrison's can be found in the literature, 
but experience has shown that analytic
modeling of the distance dependence is rarely followed in metals. Later 
Harrison included additional conduction states and concluded that the
hopping parameters for $sp^3$ compounds were better described by a $1/R^2$
dependence, albeit with a numerical parameter specific to the various
hopping parameters.\cite{Harrison1981}  As a detail: since
orthogonal TB is adopted, the orbitals are Wannier functions incorporating 
further neighbor orbitals to achieve orthogonality, with unpredictable distance 
and orientation dependence.  

In any case, any such
parametrization, whose derivative $dt/dR$ is to be used to provide the EPC
matrix element from changes in band energies, is basically
an uncontrolled simplification of the self-consistent DFT potential change
from linear response, unique to MgB$_2$, that is calculated
and applied  for EPC matrix elements in the current implementations of first 
principles theory. 

An aside: this model does not address EPC as such (as in Eq.~(\ref{eqn:lambda})).
What is evaluated is a FS average of electron-{\it ion} matrix elements as provided
by the tight binding Hamiltonian, with the $\omega_{k-k',\nu}^2$ factor in the 
denominator factored out into an effective mean value, left to be
estimated by external means. On the 
other hand, the severe approximation for the matrix elements renders close attention
to frequency factors unproductive. 

The very large matrix
elements of MgB$_2$ in DFT  occur because (i) the $2p$-$2p$ overlap between 
neighboring $\sigma$-bonded orbitals remains
large with small change of the separation $R$, while (ii) the change in potential due to
redistribution of charge in the bonding
region is very sensitive to atomic distortions, simplified in this model to 
parametrized variation
of $R$. This behavior in the bond charge region has been
understood conceptually for decades and has recently been calculated properly, but improved
understanding remains a topic for future study. 

The deformation potential approach, which connects EPC strength 
of a specific phonon displacement to
the shift of Fermi level eigenvalues and underlies expressions given by Yu2023, 
established that the splitting of the $k=0$ $\sigma$ band eigenvalues (``deformation 
potential'') for $\Gamma$ point 
due to bond stretch phonons is 13 eV/\AA -- an unprecedentedly large metallic band
shift due to EPC.\cite{An2001} MgB$_2$ is, and remains, a unique case of EPC
of a phonon coupled to FS states. 

Another feature of the model is that this $t(R)$ form
neglects bond-bending changes in the matrix elements, which physically contribute to strong
bond-bending forces. Such forces are known to be large in strongly bonded compounds such
as diamond, graphite, and MgB$_2$, and smaller but still an essential amount in Si
and Ge (to stabilize the diamond structure). 
The $\Gamma$ point $E_g$ bond-stretch modes of MgB$_2$ include two
rotating B-B bonds in addition to the B-B bond stretch, and the change in density, and
resulting potential in EPC, extends to bonding rearrangements that vary 
little with $R$.  

Finally, we return to the $\sigma$ band structure, which has a maximum at 
$\Gamma$ and downward dispersion with
increasing $k$, that accommodates holes. This band structure is 
common for quasi-2D (or strictly
2D) materials, even those with a valence band structure like MgB$_2$ but with weak EPC. The 
EPC matrix elements in MgB$_2$ are unique -- extremely strong for a small minority of
modes, weak for the remainder --  producing the premier conventional superconductor
at ambient pressure. The EPC matrix elements obtained from this TB model ($dt/dR$) are 
generic and ordinary, without accounting for the underlying mechanism. 

\section{Summary} 
The strength of the electron-phonon coupling, in the
title of this manuscript, is not manifest in the tight binding model.
Some of the points of this Comment, intended to
delineate differences between first principles theory and the
Yu2023 model as applied to the unique superconductor MgB$_2$, 
should be emphasized. \\
\vskip 1mm \noindent
$\bullet$ First principles theory agrees with, and in several cases predicted,
many properties of MgB$_2$: band structure and FSs; phonon frequencies
and linewidths that provide evidence of the giant Kohn anomaly that
would never have been suspected of a `simple' $s$-$p$ metal;
two-band behavior in $\lambda$ and the order parameter; T$_c$ (many
papers), which requires inclusion of the two-band character 
and large coupling effects (anharmonicity; non-linear EP coupling) 
to obtain a theoretically complete value of T$_c$ and to account
for the unusual two-gap shape of $C_v(T)$;\cite{SCDFT} separate Mg and B isotope shifts. 
A complete list would include  more properties.\\
\vskip 1mm \noindent
$\bullet$ The Yu2023 model is based solely on a generic tight binding form of 
Hamiltonian $h(\{t_j(R)\})$ with an $sp^2$-$p_z$ band
structure for a quasi-2D material. MgB$_2$ is exceptional, and the physics 
described in Sec. II 
is neglected in this model, so extrapolations from the model should be
scrutinized.\\
\vskip 1mm \noindent
$\bullet$ A central feature of the model is the Gaussian approximation for an
analytic form of $t(R)$. This approximation [addressed above, and see
below Eq. (A17)] ``allows us to define the energetic and geometric
parts of the EPC.'' The underlying aspects of
this separation and the later identification of the topological
part appear to be an outgrowth of the Gaussian approximation
rather than the physics of MgB$_2$.\\
\vskip 1mm \noindent
$\bullet$ Electron-ion matrix elements are obtained from $\nabla_R h$,
{\it i.e.} $\nabla_R t(R)$, and will be of
generic magnitude, whereas $\sigma$-band matrix elements in MgB$_2$,
reflected in the DFT-based $\Gamma$-point band splitting evident in the
corresponding deformation potential,\cite{An2001} are uniquely large
for a metal. The measured ratio\cite{Shukla2003} of $\gamma/\omega$ for the renormalized
$E_g$ modes is 25 meV/60 meV = 5/12; these modes (whose linewidth
includes anharmonic and non-linear coupling contributions) exist for only 2-3
vibrations before decay. None of this -- the strong B-B bonding and 
large density response to ionic motion --
can be addressed in a tight binding model of the electron bands without
phonons. (DFT or experimental
values are incorporated when needed.)\\
\vskip 1mm \noindent
$\bullet$ There is no specific treatment of phonons -- nothing specific in the
model emphasizing the essence of the 3\% of very strongly coupled
phonons versus the remaining 97\%.  The average of 
the electron-ion quantity $\Gamma_{n,n'}(k,k')$ seems to incur no
unusual physics, as phonon-related effect have been neglected. \\
\vskip 1mm \noindent
$\bullet$ The ``deep understanding'' of EPC stressed in the Abstract and
the Discussion, and the implication that existing theory is missing
something very basic, is unjustified, being based on a model that 
neglects most of the fundamental physics of EPC in MgB$_2$. The 
assertion that the topological aspect ``might favor superconductivity
with relatively high critical temperature'' is perplexing. 

\vskip 2mm 
To summarize more briefly: the electronic structure emphasized in this model is of
a generic, parametrized type, which includes no hint of the colossal
impact of the cylindrical Fermi surfaces and the uniquely large EPC
matrix elements involving the electron states on the Fermi surface
and the scattering processes across these Fermi surfaces. It would
require a much more detailed treatment to uncover additional ``deep
understanding,'' if there is more to be understood.

\vskip 2mm \noindent
{\it Acknowledgments.} I acknowledge communication with, and some clarification
 from, the authors of Ref.~[\onlinecite{Yu2023}].

\end{document}